\documentclass[aps,prd,preprint,groupedaddress]{revtex4}
\usepackage{amssymb}
\def\be{\begin{equation}}
\def\ee{\end{equation}}
\def\bes{\begin{eqnarray}}
\def\ees{\end{eqnarray}}
\def\p{\partial}

\def\half{{\textstyle\frac{1}{2}}}

\catcode`\@=11
\def\@citex[#1]#2{%
\if@filesw \immediate \write \@auxout {\string \citation {#2}}\fi
\@tempcntb\m@ne \let\@h@ld\relax \def\@citea{}%
\@cite{%
  \@for \@citeb:=#2\do {%
    \@ifundefined {b@\@citeb}%
      {\@h@ld\@citea\@tempcntb\m@ne{\bf ?}%
      \@warning {Citation `\@citeb ' on page \thepage \space undefined}}%
      {\@tempcnta\@tempcntb \advance\@tempcnta\@ne%
      \@tempcntb\number\csname b@\@citeb \endcsname \relax%
      \ifnum\@tempcnta=\@tempcntb %Number follows previous--hold on to it
        \ifx\@h@ld\relax%
          \edef \@h@ld{\@citea\csname b@\@citeb\endcsname}%
        \else%

          \edef\@h@ld{\ifmmode{-}\else--\fi\csname b@\@citeb\endcsname}%
        \fi%
      \else%   %  non-successor--dump what's held and do this one
        \@h@ld\@citea\csname b@\@citeb \endcsname%
        \let\@h@ld\relax%
      \fi}%
    \def\@citea{,\penalty\@highpenalty\,}%
  }\@h@ld
}{#1}}

\def\@citeb#1#2{{[#1]\if@tempswa , #2\fi}}
\def\@citeu#1#2{{$^{#1}$\if@tempswa , #2\fi }}
\def\@citep#1#2{{#1\if@tempswa , #2\fi}}

\def\bcites{         % cite with []'s
        \catcode`\@=11
        \let\@cite=\@citeb
        \catcode`\@=12
}

\def\upcites{         % cite with exponents
        \catcode`\@=11
        \let\@cite=\@citeu
        \catcode`\@=12
}

\def\plaincites{      % cite without brackets
        \catcode`\@=11
        \let\@cite=\@citep
        \catcode`\@=12
}

\newcount\hour
\newcount\minute
\newtoks\amorpm
\hour=\time\divide\hour by 60
\minute=\time{\multiply\hour by 60 \global\advance\minute by-\hour}
\edef\standardtime{{\ifnum\hour<12 \global\amorpm={am}%
        \else\global\amorpm={pm}\advance\hour by-12 \fi
        \ifnum\hour=0 \hour=12 \fi
        \number\hour:\ifnum\minute<10 0\fi\number\minute\the\amorpm}}
\edef\militarytime{\number\hour:\ifnum\minute<10 0\fi\number\minute}

\def\draftlabel#1{{\@bsphack\if@filesw {\let\thepage\relax
   \xdef\@gtempa{\write\@auxout{\string
      \newlabel{#1}{{\@currentlabel}{\thepage}}}}}\@gtempa
   \if@nobreak \ifvmode\nobreak\fi\fi\fi\@esphack}
        \gdef\@eqnlabel{#1}}
\def\@eqnlabel{}
\def\@vacuum{}
\def\marginnote#1{}

\begin{document}

%\hfill CERN-TH/2001-310

\preprint{UTHET-02-0801}

%\hfill {\tt hep-th/0111085} 
%\vspace{-0.2cm}

%\begin{center}
%\Large
\title{Fat Branes in Infinite-Volume Extra Space}
\thanks{Research 
supported in part by the DoE under grant DE-FG05-91ER40627.}

%\normalsize
\author{Chad Middleton}
\email{cmiddle1@utk.edu}
\author{George Siopsis}
\email{gsiopsis@utk.edu}
\affiliation{Department of Physics and Astronomy, \\
The University of Tennessee, Knoxville \\
TN 37996 - 1200, USA.}
%\end{center}
\date{April 2003}
\begin{abstract}
We study branes residing in infinite volume space and of finite extent in the transverse directions.
We calculate the graviton propagator in the harmonic gauge both inside and outside the brane and discuss its
dependence on the thickness of the brane. Our treatment includes the full tensor structure
of the propagator. We obtain two infinite towers of massive modes and
tachyonic ghosts. In the thin-brane limit,
we recover four-dimensional Einstein gravity.
We compare our results to similar recent results by Dubovsky and Rubakov.
% We find that this model gives rise to not only a massless propagator corresponding to 4D General Relativity everywhere on the brane, but also to two additional contributions which are shown to correspond to an infinite number of massive gravitons and tachyonic ghosts with a discrete mass spectrum whose contribution can be made vanishingly small in the thin-brane limit.  We find that in the small four-momentum limit, the terms combine to yield a graviton propagator which exhibits $D$ dimensional behavior both in its tensor structure and momentum dependence. This differs drastically from the large four-momentum limit, where the contributions from the massive graviton and tachyon damp out and one is left with standard 4D Einsteinian gravity.

\end{abstract}
\maketitle
%\newpage

\section{Introduction}

The weakness of the gravitational force has been successfully explained by
postulating the existence of extra dimensions~\cite{bib1,bib2,bib3,bib4,bib5}.
The effect of the extra dimensions is a high-energy modification of Newton's
Law of gravity due to the tower of Kaluza-Klein modes. When the extra dimensions are of
infinite volume, light Kaluza-Klein modes may dominate even at low energies~\cite{bib6,bib7,bib8}.
Thus, unlike with finite-volume extra space, Newton's Law is modified at astronomically large distances~\cite{bib5a,bib5b,bib5c,bib5d,bib5e,bib5f,bib5g,bib5h,bib5i}.
Dvali and Gabadadze~\cite{bib7} showed that this is not the case if the infinite space in which the brane lives has dimension $D>5$.
They studied a three-brane of the $\delta$-function type and showed that the
graviton propagator has a four-dimensional momentum dependence on the brane
even at low energies. This feature is expected to persist if the brane is
of finite thickness (``fat'') in the transverse directions for phenomenologically relevant values of the momentum. For extremely low energies, a fat brane
should lead to a higher-dimensional behavior of the propagator.
This was discussed qualitatively in~\cite{bib7}.

Here, we present a quantitative study of a fat brane in infinite volume extra space.
By linearizing gravity in the harmonic gauge, we arrive at an explicit expression for
the graviton propagator. First, we obtain the propagator for the trace of the
metric field over the transverse directions. The trace is a scalar field from the four-dimensional brane point of view. 
 This scalar then contributes to the four-dimensional graviton propagator as a source, in addition to the matter
fields. This complicates the tensor structure of the graviton propagator which becomes momentum dependent.  We explicitly obtain the solution for the graviton propagator and analyze its momentum dependence and pole structure.
We find two infinite towers of massive modes and
tachyonic ghosts.
%The contribution of the two towers vanishes i
In the thin-brane limit we
recover four-dimensional Einstein gravity on the brane.
% , which has a contribution from three terms.  The first term corresponds to that of 4D General Relativity everywhere on the brane with two additional terms which correspond to an infinite number of massive gravitons and tachyonic ghosts whose contributions vanish in the thin-brane limit.   It is found that the pole corresponding to the massless propagator of 4D Einstein gravity is independent of the bulk coordinates on the brane.  The momentum dependence of the propagator is then analyzed and we demonstate how the graviton propagator changes from one of 4D behavior to that of $D$ dimensional behavior.

Our discussion is organized as follows.
In section~\ref{sec2}, a brane-bulk action is considered which is similar to 
the action of ref.~\cite{bib7} but generalized to allow for a brane of finite 
thickness in the bulk.
In section~\ref{sec3}, we solve the linearized Einstein field equations in the harmonic gauge and
%obtain an equation for the graviton propagator.  To solve this equation, we first solve for two scalar Green functions which correspond to the trace of the worldvolume components of the Einstein equations and to a scalar propagator of the analogous scalar action.  Using these results, we then 
obtain an explicit expression for the graviton propagator.
In section~\ref{sec4}, we analyze the pole structure of the graviton propagator
and compare our results to the Dubovsky-Rubakov model~\cite{bibVVVV}.
% for a comparison of the results of our model with those of this alternative model   
In section~\ref{sec5}, we analyze the momentum dependence of the graviton
propagator both on the brane and in the bulk.
%find a critical momentum and examine our solution in the small and large momentum regime which demonstrates the momentum dependence of the graviton propagators tensor structure.  
We finally conclude with a summary of our results in~\ref{sec6}.

% which collapses to the DGP solution when 

%the brane approaches the delta-function type brane.  
%The solution is then studied as a function of momentum including its full
%tensor structure.

\section{\label{sec2} $D$-Dimensional Fat Brane Model}

We are interested in the dynamics of a 3-brane in a $D$-dimensional 
infinite
space. The 3-brane is allowed to have finite thickness in the 
bulk-space with extent governed by the density function $\sigma_\Lambda (y)$.
The action is similar to the one discussed in ref.~\cite{bib7},
\be\label{eq2}
S=M^{D-2}\int d^4xd^{D-4}y\sqrt{-g}\; \mathcal{R}^{(D)} + \overline 
M^2\int d^4xd^{D-4}y\sqrt{-\overline g}\; \sigma_\Lambda 
(y)\overline\mathcal{R}^{(4)} + S_{matter}
\ee
where $g_{AB}$ is the $D$-dimensional metric which generates the 
$D$-dimensional Ricci
scalar $\mathcal{R}^{(D)}$, whereas $\overline\mathcal{R}^{(4)}$ is generated
by the four-dimensional metric $\overline g_{\mu\nu}$ which is the induced
metric on the slice $\vec y=$~const..
Capital Latin indices run over $D$-dimensional space-time 
$(A,B=0,1,2,...,D)$, Greek indices run over the four-dimensional brane worldvolume 
spanned by coordinates $x^\mu$ $(\mu = 0,1,2,3)$ and lowercase Latin 
indices run over the extra space spanned by $y_m$ $(m=4,5,...,D)$.
$M$ is the $D$-dimensional Plank mass. $S_{matter}$ is the (unspecified)
matter action giving rise to the fat brane configuration.
We set $y = |\vec y| = \sqrt{y_1^2+y_2^2+...+y_{D-4}^2}$ and 
$\sigma_\Lambda (y)$ is
a smooth function of width $1/\Lambda$ approximating a $\delta$-function. 
The mass scale $\overline M$ is related to the four-dimensional Newton 
constant.  In general, $\overline M$ will depend on $M$, but here they 
will be treated as independent scales.

For explicit calculations, we will choose a step-function form of the
density $\sigma_\Lambda$,
\be\label{eqstep}
\sigma_\Lambda(y) = \frac{(D-4)\Lambda^{D-4}}{\omega_{D-4}} \; 
\Theta(1/\Lambda- y)\ee
where $\omega_n$ is the surface area of the unit $n$-dimensional 
sphere.
% and $N=D-4$ the number of bulk dimensions.
The careful reader may wish to 
smoothen the step-function first and then take the limit in which 
$\sigma_\Lambda$ becomes discontinuous.
Our results are not altered.
In the limit 
$\Lambda\to\infty$,
the density $\sigma_\Lambda$ approaches a $\delta$-function:
$\sigma_\Lambda(y)\to \delta^{D-4}(\vec y)$. This continuous 
distribution of the 3-brane
may be thought of as the continuous limit of a discrete set of the 
four-dimensional
hypersurfaces (infinitely thin 3-branes) discussed in~\cite{bib7}.
%We now proceed with the derivation of t

The Einstein field equations are
\be\label{eqein}
M^{D-2} G_{AB}^{(D)} (x^\mu , y^m) + \overline M^2 \sigma_\Lambda (y)
\overline G_{AB}^{(4)} (x^\mu, y^m) = T_{AB} (x^\mu, y^m)\ \ ,
\ee
where $G_{AB}^{(D)}$ is the $D$-dimensional Einstein tensor and $\overline G_{AB}^{(4)}$
only has brane worldvolume components.
%calculating the graviton propagator.
% \be
% S_{Brane}=\overline{M}^2\sum_{i=1}^N\; \int d^4x\sqrt{-\overline{g}_i(x,y_i)}\;\overline{\mathcal{R}}_{(4)}(x) 
% \ee
Expanding around a flat background,
\be g_{AB} = \eta_{AB} + h_{AB}\ee
the first-order Einstein equations are as follows.
The transverse components give
\be\label{eq8a}
2\p^A\p^n h_{An}-\p_n\p^n h_A^A-\p_A\p^Ah_n^n
= (D-4)(\p^C\p^D h_{CD}-\p_C\p^Ch_D^D)
\ee
The mixed components give
\be\label{eq8b}
\p_A\p^A h_{\alpha n} = \p_\alpha\p^A h_{A n}+\p_n\p^A h_{A 
\alpha}-\p_\alpha\p_n h_A^A
\ee
and the brane worldvolume components imply
\bes\label{eq8c}
M^{D-2}(\p_\alpha\p^A h_{\beta A}+\p_\beta\p^A h_{\alpha A}-\p_A\p^A 
h_{\alpha\beta}-\p_\alpha\p_\beta h_A^A
&-&\eta_{\alpha\beta}\left(
\p^A\p^B h_{A B}-\p_B\p^B h_A^A\right)\;)\nonumber\\
+ \overline{M}^2\sigma_\Lambda(y)\;(\p_\alpha\p^\nu 
h_{\beta\nu}+\p_\beta\p^\nu h_{\alpha\nu}-\p_\nu\p^\nu h_{\beta\alpha}-\p_\alpha\p_\beta 
h_\nu^\nu
&-&\eta_{\alpha\beta}\left( \p^\mu\p^\nu h_{\mu\nu}-\p_\mu\p^\mu 
h_\nu^\nu
\right)\;)\nonumber\\
&=& T_{\alpha\beta} (x^\mu,\vec y)
\ees
where we have chosen a matter source described by the stress-energy tensor $T_{\mu\nu}$ 
whose transverse components vanish 
($T_{mn} = T_{\mu n} = 0$).  Indices are raised and lowered by the flat metric 
tensor
$\eta_{AB}$.
% We now examine the Einstein equations in the brane and bulk when $D\geq 
% 5$.

To solve the field equations, we shall choose the harmonic gauge,
\be\label{eqhar}
\p^A h_{AB}={1\over 2}\p_B h_A^A \ \ .
\ee
We obtain from eqs.~(\ref{eq8a}) and (\ref{eq8b}),
%it can be shown that the $\{\mu n\}$ and $\{nm\}$ components of 
%Einstein's equations yield,
respectively,
\bes
(6-D) \p_A\p^A h_n^n &=& (D-4)\p_A\p^A h_\mu^\mu\label{eq2b}\\
\p_A \p^A h_{m\alpha} &=& 0\label{eq2a}
\ees
so we may set
\bes
h_{m\alpha} &=& 0\label{eq22a}\\
(D-6) h_n^n + (D-4)h_\mu^\mu &=& 0\label{eq22b}
\ees
Then the brane worldvolume components of the Einstein equations can be written 
in the following form:
\bes\label{eq3}
-M^{D-2}\p_A\p^A \left(h_{\alpha\beta}-\half \eta_{\alpha\beta}
 h_B^B \right)&+&\overline M^2\sigma_\Lambda\;\left(-\p_\nu\p^\nu 
h_{\alpha\beta}+\p_\alpha\p_\beta h_n^n-\half \eta_{\alpha\beta}\; \p^\mu\p_\mu (h_n^n- 
h_\nu^\nu) \right)\nonumber\\
&=& T_{\alpha\beta} (x^\mu,\vec y)
\ees
%where we used~(\ref{eq2a}).
Performing a Fourier transform in the brane worldvolume coordinates $x^\mu$ and
multiplying by an arbitrary
conserved stress-energy tensor ${T'}^{\alpha\beta}$, which for simplicity is
assumed to have no $\vec y$-dependence, 
we
obtain
\bes\label{eq3T11}
\left(M^{D-2}(p^2 +\p_n \p^n) + \overline {M}^2\sigma_\Lambda\; p^2 
\right) \tilde h_{\alpha\beta}{\tilde T}^{\prime\alpha\beta}
&=& \tilde T_{\alpha\beta}(p^\mu,\vec y) {\tilde T}^{\prime\alpha\beta}\nonumber\\
&+& \half {\tilde T}_\mu^{\prime\mu} \left( \overline{M}^2\sigma_\Lambda\; 
p^2\; (\tilde h_\nu^\nu - \tilde h_n^n)
+ M^{D-2} (p^2+\p_n \p^n) \tilde h_A^A\right)
\ees
%where the term $\p_\alpha\p_\beta h_n^n$ vanishes when convoluted with 
%the conserved energy-momentum tensor.
where the Fourier transformed, $D$-dimensional d'Alembertian is
$\p_A\p^A=-\p_n\p^n-p^2$
with
$p^2=p_0^2-\vec{p}^{\, 2}$ the worldvolume Minkowski four-momentum.
In the next section, we shall solve this equation for the graviton propagator.
% To solve this equation, we will first need to find $\tilde h_\nu^\nu$ and 
%$\tilde h_n^n$.

\section{\label{sec3} Graviton Propagator}

In general, the spread functions of the brane and the matter source are different.
However, it was argued by Dvali, {\em et al.}~\cite{dghs} that the two spreads
coincide at lowest order with correction terms suppressed by factors
$o(M/\overline{M})$.
We shall therefore adopt a source stress-energy tensor of the form
%proceed under this argument and set 
\be\label{eqTab} T_{\alpha\beta}(x^\mu,\vec y)= T_{\alpha\beta}(x^\mu)\;\sigma_\Lambda(y)\ee
in the explicit calculation of the tensor structure and momentum dependence of the graviton propagator.
%  We then compare the results of our model with that of Dubovsky et. al~\cite{bibVVVV}.
%
%
%
%
%
%
%\subsection{Fat Matter Source}\label{eqSP}
%
%We continue with solving the Einstein equations when the source describing the fat brane is equivalent to that of the matter source.
Taking the
trace of eq.~(\ref{eq3}), we obtain
\be\label{eq33}
-\frac{(D-2)}{(D-4)}\; M^{D-2} (p^2+\p_s\p^s) \tilde h_n^n
+ \frac{2(D-5)}{(D-4)}\; \overline{M}^2\sigma_\Lambda\;
p^2\; \tilde h_n^n
= \tilde T_\alpha^\alpha \;\sigma_\Lambda(y)
\ee
where we used eq.~(\ref{eq22b}) to express $\tilde h_\mu^\mu$ in terms 
of $\tilde h_n^n$. This is an equation for the field $\tilde h_n^n$ (trace over transverse directions of the metric field), which
is a scalar from a four-dimensional point of view.
The solution is obtained on the brane and in the bulk in terms of the
Green function to the wave equation,
\be\label{eq19}
\left( M^{D-2}\; (p^2+\p_s\p^s)- (\lambda-1)\;\overline{M}^2 p^2
\sigma_\Lambda \right) \mathcal{G}_\lambda (p, y) = \sigma_\Lambda(y)
\ee
as
%where the trace $\tilde{h}_n^n$ is proportional to the average over the source of the Green function,
\be\label{eqtr} \tilde h_n^n(p,y) = -\frac{(D-4)}{(D-2)}\; \tilde{T}_\alpha^\alpha \;\mathcal{G}_\lambda(p,y)\quad,\quad \lambda = \frac{3(D-4)}{(D-2)}
\ee
After some algebra, we obtain a spherically symmetric solution
expressed in terms of Bessel functions as
\be\label{eqGl}
\mathcal{G}_\lambda(p,y)=-\frac{1}{\overline M^2 p^2(\lambda-1)}\left( 1
- \frac{1}{\mathcal{B}_\lambda}
\left(1\over y\Lambda\right)^{(D-6)/2} \; 
H^{(1)}_{(D-4)/2}(p/\Lambda)I_{(D-6)/2}(k_\lambda py)\right)
\ee
inside the brane ($y\le 1/\Lambda$), and
\be\label{eqGla}
\mathcal{G}_\lambda(p,y)=-\frac{1}{\overline M^2 p^2(\lambda-1)}\;
\frac{k_\lambda}{B_\lambda}\left( \frac{1}{y\Lambda}\right)^{(D-6)/2}\; I_{(D-4)/2}(k_\lambda p/\Lambda)
H^{(1)}_{(D-6)/2}(py)
\ee
in the bulk ($y>\Lambda$), where
\be\label{eqGlamAB}
\mathcal{B}_\lambda
=k_\lambda I_{(D-4)/2}(k_\lambda p/\Lambda)H^{(1)}_{(D-6)/2}(p/\Lambda)+I_{(D-6)/2}(k_\lambda p/\Lambda) H^{(1)}_{(D-4)/2}(p/\Lambda)\ee
and we have introduced the constant $k_\lambda$ given by
\be\label{eqklam2} k_\lambda^2=(\lambda-1)\; \frac{(D-4)\Lambda^{D-4}\; 
\overline{M}^2}{\omega_{D-4}\; M^{D-2}}-1 \approx (\lambda-1)\; \frac{(D-4)\Lambda^{D-4}\; 
\overline{M}^2}{\omega_{D-4}\; M^{D-2}}\ee
To obtain the graviton propagator, we will also need the Green function which is
the solution to eq.~(\ref{eq19}) when $\lambda=0$.
Notice that when $\lambda=0$, eq.~(\ref{eq19}) turns into the wave
equation for a scalar field in the thin-brane limit.
It is derivable from a scalar field action.
% corresponds to that of a scalar model derivable from a scalar action.
Explicitly,
%
% We similarly obtain
%
%\bes\label{eqG0a}
%\mathcal{G}^{\mathrm{in}}_0(p,y,y')
%&=&-\frac{i\pi}{4M^{D-2}}\;{J_{(D-6)/2}(k_0 py_<)\over (yy')^{(D-6)/2}}\nonumber\\
%&\times&{1\over \mathcal{B}} \; \left(\mathcal{A}^{(2)} H^{(1)}_{(D-6)/2}(k_0py_>) - \mathcal{A}^{(1)}\; H^{(2)}_{(D-6)/2}(k_0py_>)\right)
%\ees
%\bes\label{eqG0b}
%\mathcal{G}^{\mathrm{out}}_0(p,y,y')
%=-\frac{1}{M^{D-2}}\;
%\left({1\over yy'}\right)^{(D-6)/2}\; \frac{\Lambda 
%}{p\mathcal{B}} J_{(D-6)/2}(k_0py') H^{(1)}_{(D-6)/2}(py)
%\ees
%where
%\bes\label{eqG0A}
%\mathcal{B}
%&=&k_0 H^{(1)}_{(D-6)/2}(p/\Lambda){J_{(D-4)/2}(k_0p/\Lambda)}-
% H^{(1)}_{(D-4)/2}(p/\Lambda) {J_{(D-6)/2}(k_0p/\Lambda)}\nonumber\\
%\mathcal{A}^{(1)}
%&=& k_0 H^{(1)}_{(D-4)/2}(k_0p/\Lambda)H^{(1)}_{(D-6)/2}(p/\Lambda)
%- H^{(1)}_{(D-6)/2}(k_0p/\Lambda) H^{(1)}_{(D-4)/2}(p/\Lambda)\nonumber\\
%\mathcal{A}^{(2)}
%&=& k_0 H^{(2)}_{(D-4)/2}(k_0p/\Lambda)H^{(1)}_{(D-6)/2}(p/\Lambda)
%- H^{(2)}_{(D-6)/2}(k_0p/\Lambda) H^{(1)}_{(D-4)/2}(p/\Lambda)
%\ees
%and
%\be
%  \label{eqk2} k^2=\;1+ \frac{N\Lambda^{D-4}\;
%  \overline{M}^2}{\omega_{D-4}\; M^{D-2}} \approx \frac{N\Lambda^{D-4}\;
%  \overline{M}^2}{\omega_{D-4}\; M^{D-2}}
%\ee
%
%We can now likewise obtain $\mathcal{G}_0(p,y)$ from eq.~(\ref{eqGpy1}) 
%
\be\label{eqG0}
\mathcal{G}_0(p,y)=\frac{1}{\overline M^2 p^2}\left( 1
+ \frac{1}{\mathcal{B}}
\left(1\over y\Lambda\right)^{(D-6)/2} \; 
H_{(D-4)/2}^{(1)}(p/\Lambda)J_{(D-6)/2}(\kappa py)\right)
\ee
inside the brane ($y\le 1/\Lambda$), and
\be\label{eqG0b}
\mathcal{G}_0(p,y)=-\frac{1}{\overline M^2 p^2}\;
\frac{\kappa}{\mathcal{B}}\; \left(\frac{1}{y\Lambda}\right)^{(D-6)/2}\; J_{(D-4)/2}(\kappa p/\Lambda)
H^{(1)}_{(D-6)/2}(py)
\ee
in the bulk ($y>1/\Lambda$), where
\be\label{eqG0A}
\mathcal{B}
=\kappa H^{(1)}_{(D-6)/2}(p/\Lambda){J_{(D-4)/2}(\kappa p/\Lambda)}-
 H^{(1)}_{(D-4)/2}(p/\Lambda) {J_{(D-6)/2}(\kappa p/\Lambda)}\ee
and (see eq.~(\ref{eqklam2}))
\be
  \label{eqk2} \kappa^2= -k_0^2 = \;1+ \frac{(D-4)\Lambda^{D-4}\;
  \overline{M}^2}{\omega_{D-4}\; M^{D-2}} \approx \frac{(D-4)\Lambda^{D-4}\;
  \overline{M}^2}{\omega_{D-4}\; M^{D-2}}
\ee
Notice that inside the brane, $\mathcal{G}_0(p,y)$ oscillates rapidly over the
transverse width of the brane.

We are now ready to deduce the full graviton propagator.
To this end, let us massage eq.~(\ref{eq3T11})
into the form
$$
\left(M^{D-2}(p^2+\p_n\p^n) + \overline {M}^2p^2\sigma_\Lambda(y)\;
\right)\left\{\tilde{h}_{\alpha\beta}(p,y)\tilde{T}^{\prime\alpha\beta}+\frac{(D-5)}{3(D-4)}\tilde{h}_n^n(p,y)\tilde{T}_\nu^{\prime\nu}\right\} $$
\be\label{eq3T1}
=\left\{\tilde{T}_{\alpha\beta}\tilde{T}^{\prime\alpha\beta}-\frac{1}{3}\tilde{T}_\mu^\mu\tilde{T}_\nu^{\prime\nu}\right\}\sigma_\Lambda(y)
\ee
The solution for the graviton propagator is readily obtained in terms of the scalar propagators,
\be\label{eqtra}
\tilde{h}_{\alpha\beta}(p,y)\tilde{T}^{\prime\alpha\beta}=
\left\{\tilde{T}_{\alpha\beta}\tilde{T}^{\prime\alpha\beta}-
\frac{1}{3}\tilde{T}_\mu^\mu\tilde{T}_\nu^{\prime\nu}\right\}\mathcal{G}_0(p,y)+
\frac{(D-5)}{3(D-2)}\tilde{T}_\mu^\mu\tilde{T}_\nu^{\prime\nu}\mathcal{G}_\lambda(p,y)
\ee
where we used eqs.~(\ref{eq19}) and (\ref{eqtr}).

\section{\label{sec4} Poles of the graviton propagator}

Next, we analyze the pole structure of the graviton propagator.
We then compare the results of our model with that of Dubovsky and
Rubakov~\cite{bibVVVV}.

\subsection{Our model}

Using the expressions~(\ref{eqG0}) for $\mathcal{G}_0(p,y)$ and~(\ref{eqGl})
for $\mathcal{G}_\lambda(p,y)$,
the graviton propagator~(\ref{eqtra}) inside the brane ($y\le 1/\Lambda$)
can be written in the form
$$\tilde h_{\alpha\beta} (p,y){\tilde T}^{\prime\alpha\beta} = \left\{ \tilde T_{\alpha\beta}{\tilde T}^{\prime\alpha\beta} - 
\frac{1}{2}
\tilde T_\alpha^\alpha \tilde T_\beta^{\prime\beta}\right\} \frac{1}{\overline M^2p^2}\;\;\;+\;\;\left(\frac{1}{y\Lambda}\right)^{(D-6)/2}H^{(1)}_{(D-4)/2}(p/\Lambda)$$
\bes\label{prpgy}
&\times&\frac{1}{\overline M^2p^2}\left[\left\{ \tilde T_{\alpha\beta}{\tilde T}^{\prime\alpha\beta} - 
\frac{1}{3}
\tilde T_\alpha^\alpha \tilde T_\beta^{\prime\beta}\right\} \frac{1}{\mathcal B}J_{(D-6)/2}(\kappa py) + 
\frac{1}{6}
\tilde T_\alpha^\alpha \tilde T_\beta^{\prime\beta}\frac{1}{\mathcal B_\lambda}I_{(D-6)/2}(k_\lambda py)\right]
\ees
For convenience, we have separated the term that corresponds to the tensor structure and
momentum dependence of the four-dimensional graviton propagator.
% on the brane with no dependence on the bulk coordinates.
To study the pole structure, we shall introduce the average value of the
graviton propagator over the transverse directions of the
brane
(see~\cite{bibkir} for problems associated with the
definition of observables on the brane)
defined by
%The average value over transverse directions of the graviton propagator can be defined in the same manner as the average values of the Green functions from the previous section.
\be\label{eqaveb}
\tilde h^{\mathrm{Brane}}_{\alpha\beta} (p)=\int d^{D-4}y\sigma_\Lambda(y)\tilde{h}_{\alpha\beta}(p,y)
\ee
%which is written in terms of the average values of the scalar propagators
%\bes\label{eqhT}
%&&\tilde h^{\mathrm{Brane}}_{\alpha\beta} (p){\tilde T}^{\prime\alpha\beta} \left\{ \tilde T_{\alpha\beta}{\tilde T}^{\prime\alpha\beta} - 
%\frac{1}{3}
%\tilde T_\alpha^\alpha \tilde T_\beta^{\prime\beta}\right\} \mathcal{G}^{\mathrm{Brane}}_0(p)+ \frac{(D-5)}{3(D-2)}\;  
%\tilde{T}_\alpha^\alpha \tilde{T'}_\beta^\beta \mathcal{G}^{\mathrm{Brane}}_\lambda(p)
%\nonumber\\
%\ees
%Using eqs.~(\ref{eq22aa}) and (\ref{eq22BB}) we can write the value for the graviton propagator as 
Integrating~(\ref{prpgy}), we obtain
\bes\label{prpg}
\tilde h^{\mathrm{Brane}}_{\alpha\beta} (p){\tilde T}^{\prime\alpha\beta} &=&
\left\{ \tilde T_{\alpha\beta}{\tilde T}^{\prime\alpha\beta} - \frac{1}{2}
\tilde T_\alpha^\alpha \tilde T_\beta^{\prime\beta}\right\} \frac{1}{\overline M^2p^2}
+ \left\{ \tilde T_{\alpha\beta}{\tilde T}^{\prime\alpha\beta} - 
\frac{1}{3}
\tilde T_\alpha^\alpha \tilde T_\beta^{\prime\beta}\right\} \frac{(D-6)}{(
\overline M \kappa p^2/\Lambda)^2[1-\mu_0(\kappa p/\Lambda)]} \nonumber\\
&-&
\frac{1}{6}
\tilde T_\alpha^\alpha \tilde T_\beta^{\prime\beta}\frac{(D-6)}{(
\overline M k_\lambda p^2/\Lambda)^2[1+\mu_\lambda(k_\lambda p/\Lambda)]}
\ees
from which we may easily deduce
the pole structure of the graviton propagator inside the brane.
The above expression is valid for $D>6$ (for $D=6$, we obtain logarithmic corrections,
but the results are similar and will not be explicitly discussed here).
The functions that appear in the denominators in~(\ref{prpg}) are
\be\label{m1}
\mu_0(z)=\frac{D-6}{z}\; \frac{J_{(D-6)/2}(z)}{J_{(D-4)/2}(z)}
\ \ , \ \
\mu_\lambda(z)=\frac{D-6}{z}\; \frac{I_{(D-6)/2}(z)}{I_{(D-4)/2}(z)}
\ee
for $D>6$.
%\be
%f_0(z)= \frac{D-6}{z^2}\;
%\left( \mu_0 (z) +1 \right)
%\ \ , \ \
%f_\lambda(z)= \frac{D-6}{z^2}\;
%\left( \mu_\lambda (z) -1 \right)
%\ee
%The $p^2=0$ pole corresponds to the massless propagator of 4D Einstein gravity with the correct tensor structure in the thin-brane limit $\Lambda\to \infty$.
%This fat brane model also predicts additional poles which 
The poles of the propagator are solutions to the equations
\be \mu_0 (\kappa p/\Lambda) = 1\ \ , \ \ \mu_\lambda (k_\lambda p/\Lambda) = -1
\ee
Using~(\ref{m1}) and the Bessel function identity
\be\label{eqbes1} z J_{\nu -1} (z) + zJ_{\nu +1} (z) = 2\nu J_\nu (z)\ee
for $\nu = (D-6)/2$, it is easily shown
that the solutions to $\mu_0 (z)=1$ are the roots of $J_{\nu-1} = J_{(D-8)/2}$.
As is well-known,
there are infinitely many zeros for $\nu >0$, i.e., $D>6$,
which is the case we are considering here.
We shall denote them by $z_j$,
\be J_{(D-8)/2} (z_j) = 0 \ \ , \ \ j=1,2,\dots \ee
We therefore obtain an infinite tower of massive poles with masses given by
\be\label{eqmp1} m_j^2 = z_j^2\, \frac{\Lambda^2}{\kappa^2} \ee
Similarly, the condition $\mu_\lambda (z) = -1$, together with the Bessel
function identity
\be z I_{\nu -1} (z) - zI_{\nu +1} (z) = 2\nu I_\nu (z)\ee
and the relation $I_\nu (z) = e^{-\pi \nu i/2} J_\nu (iz)$, lead to a tower of
tachyonic poles with masses given by
\be\label{eqmp2} m_{*j}^2 = - z_j^2\, \frac{\Lambda^2}{k_\lambda^2} \ee
To obtain the behavior of the propagator near a massive pole, observe that
\be 1-\mu_0(z) = - \frac{J_{(D-8)/2} (z)}{J_{(D-4)/2} (z)} = -
\frac{J_{(D-8)/2}' (z_j)}{J_{(D-4)/2} (z_j)} \, (z-z_j) + o((z-z_j)^2) \ee
Using the Bessel function identity
\be zJ_{\nu-1}' (z) = (\nu -1) J_{\nu -1} (z) - zJ_\nu (z)\ee
together with~(\ref{eqbes1}), we deduce
\be 1-\mu_0(z) =
\frac{1}{2(D-6)} \, (z^2-z_j^2) + \dots \ee
near $z=z_j$. It follows that the graviton propagator on the brane~(\ref{prpg})
behaves as
\be\label{prpga}
\tilde h^{\mathrm{Brane}}_{\alpha\beta} (p){\tilde T}^{\prime\alpha\beta} \sim
\left\{ \tilde T_{\alpha\beta}{\tilde T}^{\prime\alpha\beta} - 
\frac{1}{3}
\tilde T_\alpha^\alpha \tilde T_\beta^{\prime\beta}\right\} \; \frac{2(D-6)^2/z_j^4}{
\overline M^2 (p^2 - m_j^2)}
\ee
near the massive pole $p^2 = m_j^2$.
Similarly, near the tachyonic pole $p^2 = m_{*j}^2$, we obtain
\be\label{prpgb}
\tilde h^{\mathrm{Brane}}_{\alpha\beta} (p){\tilde T}^{\prime\alpha\beta} \sim
- 
\frac{1}{6}
\tilde T_\alpha^\alpha \tilde T_\beta^{\prime\beta} \; \frac{2(D-6)^2/z_j^4}{
\overline M^2 (p^2 - m_{*j}^2)}
\ee
%with the functions $m^2(p)$ and $m_*^2(p)$ given by
%\bes\label{m1a}
%m^2(p)&=&\left(\frac{D-6}{2}\right)^2\left(\frac{2\Lambda}{\kappa }\right)^2\frac{J^2_{(D-6)/2}(\kappa  p/\Lambda)}{J^2_{(D-4)/2}(\kappa  p/\Lambda)}\nonumber\\
%&=&\left(\frac{\Lambda}{\kappa }\right)^2\frac{J^2_{0}(\kappa  p/\Lambda)}{J^2_1(\kappa  p/\Lambda)\mbox{ln}^2(p/2\Lambda)}\;\;\;\mbox{for}\;(D=6)
%\ees
%\bes\label{m2a}
%m_*^2(p)&=&\left({D-6\over 2}\right)^2\left(\frac{2\Lambda}{k_\lambda}\right)^2\frac{I^2_{(D-6)/2}(k_\lambda p/\Lambda)}{I^2_{(D-4)/2}(k_\lambda p/\Lambda)}\nonumber\\
%&=&\left(\frac{\Lambda}{k_\lambda}\right)^2\frac{I^2_{0}(k_\lambda p/\Lambda)}{I^2_1(k_\lambda p/\Lambda)\mbox{ln}^2(p/2\Lambda)}\;\;\;\mbox{for}\;(D=6)
%\ees
%
%which correspond to the masses of each propagator when evaluated at each particular pole.
%The functions $f_0\left(\kappa p/\Lambda\right)$ and $f_{\lambda}\left(k_\lambda p/\Lambda\right)$ are
%\bes
%f_0(\kappa p/\Lambda)&=&\left(\frac{\kappa p}{2\Lambda}\right)^{-3}\left[\frac{\Gamma({D-4\over 2})J_{(D-6)/2}(\kappa p/\Lambda)+\Gamma({D-6\over 2})\left({\kappa p/ 2\Lambda}\right)J_{(D-4)/2}(\kappa p/\Lambda)}{2\Gamma^2({D-6\over 2})J_{(D-4)/2}(\kappa p/\Lambda)/\Gamma({D-4\over 2})}\right]\nonumber\\
%\ees
%\bes
%f_\lambda(k_\lambda p/\Lambda)&=&\left(\frac{k_\lambda p}{2\Lambda}\right)^{-3}\left[\frac{\Gamma({D-4\over 2})I_{(D-6)/2}(k_\lambda p/\Lambda)-\Gamma({D-6\over 2})\left({k_\lambda p/ 2\Lambda}\right)I_{(D-4)/2}(k_\lambda p/\Lambda)}{2\Gamma^2({D-6\over 2})I_{(D-4)/2}(\kappa \lambda p/\Lambda)/\Gamma({D-4\over 2})}\right]\nonumber\\
%\ees
%
%
The minus sign of the residue of the tachyon implies that the tachyon is a ghost. 

Notice that both the massive modes~(\ref{eqmp1}) and the tachyons~(\ref{eqmp2})
are expressed in terms of the same mass scale parameter $p_c$, where
\be\label{eqpc}
p_c^2 \sim \frac{\Lambda^2}{\kappa^2} \sim \frac{\Lambda^2}{k_\lambda^2}
\sim \frac{M^{D-2}}{\overline M^2 \Lambda^{D-6}}
\ee
In the thin-brane limit ($\Lambda\to \infty$), we have $p_c\to 0$ and the
infinite twoers of massive modes and tachyons turns into continuous spectra.
The form of the propagator in this limit is easily deduced from eq.~(\ref{prpg}).
For momenta away from the critical scale ($|p|\gg p_c$), the two
terms in~(\ref{prpg}) that give rise to the massive and tachyonic poles
become vanishingly small and we are left with
%We can now examine the thin-brane limit of the spread functions of the matter source and the brane.  Taking the $\Lambda\rightarrow\infty$ limit,
% the functions $f_0(k_0p/\Lambda)$ and $f_\lambda(k_\lambda p/\Lambda)$ tend toward zero and 
\be\label{eqearl}
\tilde h_{\alpha\beta}^{\mathrm{Brane}}(p){\tilde T}^{\prime\alpha\beta}
\sim
\left\{ \tilde T_{\alpha\beta}{\tilde T}^{\prime\alpha\beta}-{1\over 
2}\tilde T_\alpha^\alpha \tilde T_\beta^{\prime\beta}\right\}{1\over 
\overline M^2 p^2}
\ee
recovering four-dimensional Einstein gravity.
% in the thin-brane limit for non-vanishing four momentum.
%
% In summary, we have obtained an expression for the graviton propagator which contains three terms consisting of the massless propagator of 4D General Relativity and two additional terms which correspond to an infinite number of massive gravitons and tachyonic ghosts with a discrete mass spectrum.  In the thin-brane limit, the contributions from the massive gravitons and tachyonic ghosts tend to zero and one is left with that of 4D Einstein gravity.

\subsection{The Dubovsky-Rubakov model}

It is interesting to note that similar results have been obtained by Dubovsky and Rubakov~\cite{bibVVVV} using a slightly different model.
In order to directly compare our results with theirs, we shall assume that the
spread function (denoted by $f^2(y)$ in~\cite{bibVVVV}) is given by eq.~(\ref{eqstep}).
Then the Einstein field equations proposed in~\cite{bibVVVV} can be written as
%We wish to compare the results of our model to those of this alternative model and examine the differences.  To briefly reiterate~\cite{bibVVVV} , they begin with the Einstein equations with an averaged brane contribution
\be\label{eqein2}
\mathcal F(\Box^{(D)})G_{AB}^{(D)}(x^\mu ,\vec y)+\overline M^2\sigma_\Lambda(y)\int d^{D-4}y'\sigma_\Lambda(y')\; \overline G_{AB}^{(4)}(x^\mu ,\vec y')=T_{AB}(x^\mu ,\vec y)
\ee
to be compared with the Einstein eq.~(\ref{eqein}) in our model.
In eq.~(\ref{eqein2}), the four-dimensional Einstein tensor only has brane
worldvolume components
(i.e., $G_{aB}^{(4)}=0$) and the form-factor $\mathcal F\approx M^{D-2}$ at low energies.
Also, the matter source on the brane will be assumed to have only space-time components
$T_{\mu\nu}$ and a spread function same as that of the brane,
\be
T_{\mu\nu}(x,y)=T_{\mu\nu}(x)\sigma_\Lambda(y)
\ee
where $T_{\mu\nu}(x)$ is conserved in the four-dimensional sense
({\em cf.}~eq.~(\ref{eqTab}) in our model).
The inverse width $\Lambda$ of the spread function is assumed to be $\Lambda\sim M$
in~\cite{bibVVVV} to be contrasted with our model in which $\Lambda\sim
\overline M$, since it coincides with the inverse width of the brane~\cite{bib7}.

Working as in section~\ref{sec2}, we linearize the Einstein equations and
obtain the graviton propagator in the form
%This yields the following for the metric perturbations
\be\label{eqpropdr}
\tilde h_{\mu\nu}(p,y) \tilde T^{\prime\mu\nu}=\frac{2}{\mathcal{C}}
\left\{ \tilde T_{\mu\nu} \tilde T^{\prime\mu\nu} - \frac{1}{3}
\tilde T_\mu^\mu \tilde T_\lambda^{\prime\lambda} \right\}
\mathcal{G}_1 (p,y) - \frac{1}{3\mathcal{C}_*} \tilde T_\mu^\mu \tilde T_\lambda^{\prime\lambda}
\mathcal{G}_1 (p,y)
\ee
%{M_{Pl}^2}\frac{D_f(p,y)}{D_{ff}(p)}\left[\frac{1}{p^2-m^2(p)}\;\left(\theta_{\mu\nu}-\frac{1}{3}\eta_{\mu\nu}\theta_\lambda^\lambda\right)-\frac{1}{6}\;\frac{1}{p^2-m_*^2(p)}\eta_{\mu\nu}\theta_\lambda^\lambda\right]
where we multiplied by the arbitrary stress-energy tensor $T_{\mu\nu}'$ to
absorb the longitudinal part which is not gauge-invariant.
It is given in terms of the Green function which satisfies eq.~(\ref{eq19}) for
$\lambda = 1$,
\be\label{eq19a}
M^{D-2}\; (p^2+\p_s\p^s)
\mathcal{G}_1 (p, y) = \sigma_\Lambda(y)
\ee
(denoted by $D_f$ in~\cite{bibVVVV}).
The denominators are
\be \mathcal{C} = 1+\overline M^2 p^2 \overline\mathcal{G}_1\ \ , \ \
\mathcal{C}_* = 1-\overline M^2 p^2 \overline\mathcal{G}_1\ee
where $\overline\mathcal{G}_1$ is the average of $\mathcal{G}_1$ over the
spread function (defined as in eq.~(\ref{eqaveb}) and denoted by $D_{ff}$ in~\cite{bibVVVV}).
Explicitly,
\be
\overline\mathcal{G}_1(p)=-\frac{\kappa^2}{\overline M^2 \Lambda^2}\, f(p/\Lambda)
\ \ , \ \ f(z) = \frac{1}{z^2}\left[1-\frac{i\pi}{2}(D-4)H^{(1)}_{(D-4)/2}(z)J_{(D-4)/2}(z)\right]
\ee
where we introduced the function $f(z)$ for convenience and the scale $\kappa$, which coincides with our earlier
definition~(\ref{eqk2}) in the large $\Lambda$ limit,
\be \kappa^2 = \frac{(D-4)\Lambda^{D-4}\overline M^2}{\omega_{D-4}M^{D-2}}\ee
The poles of the propagator~(\ref{eqpropdr}) are the zeros of $\mathcal{C}$ and $\mathcal{C}_*$.
They can easily be seen to correspond to small $z$, therefore we may approximate
$\mathcal{C} \approx 1- \kappa^2 f(0) p^2/\Lambda^2$, whose root is
\be m^2 \approx \frac{\Lambda^2}{\kappa^2 f(0)} \sim \frac{M^{D-2}}{\overline M^2 \Lambda^{D-6}} \ \ , \ee
which is a massive pole. Similarly, the root of $C_*$ is a tachyonic pole
\be m_*^2 \approx - m^2 \sim - \frac{M^{D-2}}{\overline M^2 \Lambda^{D-6}}\ee
Notice that the mass scale is similar to the mass scale of the poles in our
model~(\ref{eqpc}), although in this model only one
pair of poles is obtained instead of the infinite tower we found in our model.
This scale matches the one found in~\cite{bibVVVV} if we set $\Lambda\sim M$, in which
case $m \sim M^2/\overline M$.

\section{Momentum Dependence of the Graviton Propagator}\label{sec5}

Having understood the large $\Lambda$ limit, we now turn to a study of the
momentum dependence of the graviton propagator keeping $\Lambda$ finite.
By introducing the width $1/\Lambda$, we have added a scale to the theory
in addition to the mass scales $M$ and $\overline M$.
It follows from the explicit form of the propagator that the relevant scales
are $\Lambda$ and $\Lambda/k$, where $k\sim k_\lambda\sim \kappa$ is a dimensionless parameter given
by~(\ref{eqk2}) or (\ref{eqklam2}). Phenomenologically, one expects $\Lambda\sim
\overline M$ and $M\ll \overline M$. So we shall restrict attention to momenta
that are well below the scale $\Lambda$ ($p\ll\Lambda$).
This range is divided by the scale
given by eq.~(\ref{eqpc})
%\be p_c = \frac{\Lambda}{k}\ee
into a small momentum ($p\ll p_c$) and a large momentum ($p\gg p_c$) regime.
Qualitatively, one expects four-dimensional behavior of the graviton propagator
for large momenta and $D$-dimensional behavior for small momenta~\cite{bib7}.
We wish to study this behavior quantitatively.

For small momentum, $p\ll p_c$, we have
\be\label{eqrel}
\mathcal{G}_0(p,y)\approx \mathcal{G}_\lambda(p,y)\ee
as can easily be verified from eqs.~(\ref{eqGla}) and (\ref{eqG0b})
in the bulk and eqs.~(\ref{eqGl}) and (\ref{eqG0}) on the brane.
The resulting tensor structure of the graviton propagator~(\ref{eqtra}) is
%  the second term on the right-hand side of eq.~(\ref{eqsol})
%does not contribute to the graviton propagator. Therefore, we obtain
\be\label{eqsolap}
\tilde h_{\alpha\beta} (p;y){\tilde T}^{\prime\alpha\beta} \approx
\left\{ \tilde T_{\alpha\beta}{\tilde T}^{\prime\alpha\beta} - 
\frac{1}{(D-2)}
\tilde T_\alpha^\alpha \tilde T_\beta^{\prime\beta}\right\} \mathcal{G}_0 
(p,y)\ee
In the bulk, we deduce from~(\ref{eqG0b}),
\be \mathcal{G}_0(p;y)\sim i\left({1\over py}\right)^{(D-6)/2}H^{(1)}_{(D-6)/2}(py)\ee
which is the propagator for a $D$-dimensional scalar field.
Therefore, the graviton behaves as a $D$-dimensional field in both its
momentum dependence and its tensor structure in the bulk.
%Notice also that in the $p\to 0$ limit, we reproduce our earlier result, eq.~(\ref{eqsola}), which is valid at $p=0$.

On the brane, after averaging over its transverse width, eq.~(\ref{eqsolap})
yields in the regime $p\ll p_c$
\bes\label{eqpcs2}
\tilde{h}_{\alpha\beta}^\mathrm{Brane}(p){\tilde T}^{\prime\alpha\beta}&\sim & 
\frac{1}{\overline{M}^2\; p^2}\; \left\{ \tilde T_{\alpha\beta}{\tilde T}^{\prime\alpha\beta} - 
{1\over(D-2)}\tilde T_\alpha^\alpha {\tilde T}_\beta^{\prime\beta}\right\}\nonumber \\
&\times & \left( 1
+ \frac{1}{\mathcal{B}\; \Gamma\left({D-4\over 2}\right)}\;
\left({\kappa p\over 2\Lambda}\right)^{(D-6)/2} \; 
H^{(1)}_{(D-4)/2}(p/\Lambda)\right)\ees
where we used eqs.~(\ref{eqG0}) and (\ref{eqaveb}).
%the fact that $G_0^\mathrm{Brane}(p)\sim G_0^\mathrm{Brane}(p)$ in the low momentum limit. 
It is easy to see that the $1/p^2$ pole vanishes.
The first non-analytic term can be found from the expansion for small argument
\be H_\nu (z) = -\frac{i}{\pi}\Gamma(\nu) \left( \frac{z}{2} \right)^{-\nu} (1+\dots ) +
\frac{2i}{\pi\Gamma (\nu+1)}  \left( \frac{z}{2} \right)^\nu \ln \left(\frac{z}{2}\right) + \dots
\ee
for integer $\nu$, where the dots represent higher-order and analytic terms.
Applying this to eq.~(\ref{eqpcs2}), we obtain
\be 
\tilde{h}_{\alpha\beta}^\mathrm{Brane}(p){\tilde T}^{\prime\alpha\beta}\sim
\left\{ \tilde T_{\alpha\beta}{\tilde T}^{\prime\alpha\beta} - 
{1\over(D-2)}\tilde T_\alpha^\alpha {\tilde T}_\beta^{\prime\beta}\right\}\;
\left( \frac{p}{\Lambda} \right)^{D-6} \ln (p/\Lambda)\ee
exhibiting $D$-dimensional behavior.
Similar conclusions may be drawn for the trace $\tilde h_n^n$ in the
small momentum regime ($p\ll p_c$).

In the large momentum regime ($\Lambda \gg p\gg p_c$), the results are similar
to those in the large $\Lambda$ limit,
which we discussed in the previous section.
In this regime, the scalar Green functions are related by
\be \mathcal{G}_0 (p,y) \approx (\lambda -1) \mathcal{G}_\lambda (p,y)\ee
to be contrasted with the relation~(\ref{eqrel}) in the regime $p\ll p_c$.
Thus the tensor structure of the graviton propagator~(\ref{eqtra}) becomes
\be\label{eqsolapa}
\tilde h_{\alpha\beta} (p;y){\tilde T}^{\prime\alpha\beta} \approx
\left\{ \tilde T_{\alpha\beta}{\tilde T}^{\prime\alpha\beta} - 
\half
\tilde T_\alpha^\alpha \tilde T_\beta^{\prime\beta}\right\} \mathcal{G}_0 
(p,y)\ee
exhibiting four-dimensional behavior.
Inside the brane, we deduce from (\ref{eqG0})
\be
\tilde h_{\alpha\beta}^{\mathrm{Brane}}(p){\tilde T}^{\prime\alpha\beta}=
 \left\{ \tilde T_{\alpha\beta}{\tilde T}^{\prime\alpha\beta}-{1\over 
2}\tilde T_\alpha^\alpha \tilde T_\beta^{\prime\beta}\right\}{1\over 
\overline M^2 p^2} 
+ o \left( (p_c/p)^2 \right)
\ee
%for the matter source of finite thickness.  This corresponds to the correct 
exhibiting the distance dependence of Newtonian gravity with the tensor structure of four-dimensional Einstein gravity.
This is in agreement with our earlier conclusion~(\ref{eqearl}) in the large
$\Lambda$ limit.
In the bulk ($y> 1/\Lambda$), we deduce from~(\ref{eqG0b})
\be
\tilde h_{\alpha\beta}(p,y){\tilde T}^{\prime\alpha\beta}\approx
-\frac{i\pi}{\Gamma(\frac{D-6}{2})}\;
\frac{1}{\overline M^2 p^2}\; \left( \frac{p^2}{2\Lambda^2} \right)^{(D-6)/2}
\; \left\{ \tilde T_{\alpha\beta}{\tilde T}^{\prime\alpha\beta}-{1\over 
2}\tilde T_\alpha^\alpha \tilde T_\beta^{\prime\beta}\right\}
\; \left( \frac{1}{py} \right)^{(D-6)/2} H_{(D-6)/2}^{(1)} (py)
\ee
Therefore, the propagator vanishes in the thin brane limit ($\Lambda\to\infty$).
These results for the momentum dependence of the tensor structure of the graviton propagator are in agreement with the qualitative suggestions of~\cite{bib7}.

\section{\label{sec6} Conclusion}

The main objective of the present work was to analyze quantitatively the
gravitational effects on a brane of finite extent in the transverse directions (``fat''). We obtained an equation for the graviton propagator which we then
proceeded to solve in two steps. First, we obtained the propagator for the
trace $h_n^n$ over the transverse directions of the metric field, which is
a scalar from the four-dimensional point of view.
% This propagator exhibited an oscillatory behavior both in the bulk and on the brane.
%Thus, the limit of a $\delta$-function brane was ambiguous. However, radiative
%corrections are expected to make scalars such as $h_n^n$ heavy and therefore
%they will decouple from low energy dynamics~\cite{bib7}. 
The trace $h_n^n$ acted as an effective source term for the graviton propagator in addition to the contribution from the matter source.
This complicated the tensor structure of the graviton propagator which became momentum dependent.
We found a solution for the graviton propagator which explicitly revealed its pole structure.
%This propagator contains the massless graviton of 4D General Relativity, which has no bulk coordinate dependence, plus two additional terms which we then showed correspond to an 
We obtained infinite towers of massive gravitons and tachyonic ghosts with a discrete mass spectrum.
We found the contributions from the massive gravitons and tachyonic ghosts in the thin-brane limit
and showed that the tensor structure and distance dependence of four-dimensional
Einstein gravity is recovered in this limit.
We then analyzed the tensor structure of the momentum dependent graviton propagator for a brane of finite thickness.
In the small momentum regime, the graviton propagator exhibited a $D$-dimensional behavior,
which was in contrast to the large momentum regime (above the critical scale $p_c$ (eq.~(\ref{eqpc})) but well below the inverse brane width $\Lambda$),
where
the contributions from the massive gravitons and tachyonic ghosts conspired
to produce
a propagator on the brane whose tensor structure and distance dependence
was that of four-dimensional Einstein gravity.

%\newpage

\end{document}